\documentclass[pre,aps,epsfig,twocolumn,showpacs]{revtex4}
\usepackage{graphicx}
\begin{document}

\title{Anomalous kinetics of attractive $A+B \rightarrow 0$ reactions}

\author{Sungchul Kwon, S. Y. Yoon, and Yup Kim}
\affiliation{Department of Physics and Research Institute of Basic
Sciences, Kyung Hee University, Seoul 130-701, Korea}

\date{\today}

\begin{abstract}
We investigate the kinetics of $A+B \rightarrow 0$ reaction with
the local attractive interaction between opposite species in one
spatial dimension. The attractive interaction leads to isotropic
diffusions inside segregated single species domains, and
accelerates the reactions of opposite species at the domain
boundaries. At equal initial densities of $A$ and $B$, we
analytically and numerically show that the density of particles
($\rho$), the size of domains ($\ell$), the distance between the
closest neighbor of same species ($\ell_{AA}$), and the distance
between adjacent opposite species ($\ell_{AB}$) scale in time as
$\rho \sim t^{-1/3}$, $\ell_{AA} \sim t^{1/3}$, and $\ell \sim
\ell_{AB} \sim t^{2/3}$ respectively. These dynamical exponents
form a new universality class distinguished from the class of
uniformly driven systems of hard-core particles.
\end{abstract}

 \maketitle


The irreversible two-species annihilation reaction $A+B
\rightarrow 0$ has been intensively and widely investigated as a
basic model of various phenomena in physics \cite{first,phy},
chemistry \cite{chem}, and biology \cite{bio}.
The reaction includes two species $A$ and $B$ which are initially
distributed at random in space. The reaction of opposite species
instantaneously takes place with a rate $k$ when two particles of
opposite species encounter on the same site (generally within a
reaction radius) during the motion of particles. The reaction
forms the third inert species, which is then disregarded in the
motion of $A$ and $B$.

For the same initial density of $A$ and $B$, $\rho_A(0) = \rho_B
(0)$, the mean-field equations predict that $\rho_A$ and $\rho_B$
decay linearly in time as $(kt)^{-1}$. However it turned out that
the random fluctuation of the initial number of two species
evolves to make a segregation into $A-$ or $B-$rich area
\cite{first,monopole,kang,zumofen}. The fluctuation and
segregation develop in time, so that the reactions of two opposite
species take place only at the boundaries of two adjacent
segregated domains. As a result, in sufficiently low dimensions,
the effect of fluctuation leads to the anomalous kinetics. That
is, the evolution of the density of particles strongly depends on
fluctuations, and cannot be derived from mean-field rate
equations.

The density decay has been known to depend on the motion of the
particles and the mutual statistics of particles. For isotropic
diffusions, the density $\rho_A (t)$ scales in time $t$ as $\rho_A
(t) \sim t^{-d/4}$ in $d$-dimensions ($d \leq d_c = 4$)
\cite{monopole,kang,zumofen,namb,length,rg}. The dimension $d_c$
is the upper critical dimension above which the density decay
follows the mean-field rate equations, so $\rho(t) \sim t^{-1}$
for $d>d_c$. With the global relative drift of one species,
$\rho_A (t)$ scales as $\rho_A (t) \sim t^{-(d+1)/4}$ for $d \leq
3$ \cite{kang}. The hard-core (HC) interaction between identical
particles is irrelevant to the case of the isotropic diffusion and
the relative drift \cite{kang}. However when both species are
uniformly driven to the same direction, the HC interaction
completely changes the asymptotic scaling as $\rho_A \sim
t^{-(d+1)/6}$ for $d \leq 2$, $t^{-d/4}$ for $2 < d \leq 4$, and
finally $t^{-1}$ for $d>4$ \cite{hardcore,hardcore2}. Without the
HC interaction, $\rho_A (t)$ decays as $\rho_A \sim t^{-d/4}$ as
in the isotropic diffusion due to Galilean invariance. Recent
studies on the reaction under L\'{e}vy mixing \cite{levy} and on
scale free networks \cite{network} showed that some mixing
mechanisms that homogenizes reactants can suppress the role of the
fluctuations.

In reality in which oppositely charged particles recombinate into
inert particles such as electron-hole recombinations in
photoluminescence \cite{Wen} and particle-antiparticle reactions
in the early universe \cite{first}, the local attractive
interaction between opposite species should be much more important
than the global uniform biases. In this paper, we investigate the
kinetics of $A+B \rightarrow 0$ reaction with the attractive
interaction between opposite species in one dimension, which may
model composite systems of oppositely charged particles much more
effectively than the model with the global and uniform drift. When
a particle is surrounded by two same species neighbors such as
$BAB$, the central particle ($A$) performs random walks. If two
opposite species particles surround a particle such as $AAB$, the
central particle ($A$) is ballistically driven to its opposite
species ($B$). As a result, the attractive interaction depends on
the local configurations of adjacent particles, and accelerate the
reactions of opposite species at the boundaries of segregated
domains. However inside segregated domains, each particle has the
same neighboring species, so the motion is isotropic diffusion.
Hence the situation belongs to neither the relative drift nor the
isotropic diffusion. Due to the isotropic diffusion inside
domains, the HC interaction should be irrelevant in this case.

With the local attractive interaction in one spatial dimension, we
analytically and numerically show, regardless of the existence of
the HC interaction, that the density of particles ($\rho$), the
distance between the closest neighbor of same species
($\ell_{AA}$), the size of domains ($\ell$), and the distance
between adjacent opposite species ($\ell_{AB}$) scale in time as
$\rho \sim t^{-1/3}$, $\ell_{AA} \sim t^{1/3}$, and $\ell \sim
\ell_{AB} \sim t^{2/3}$ respectively. These dynamical exponents
form a new universality class distinguished from the class of
uniformly driven systems of hard-core particles
\cite{hardcore,hardcore2}, where the exponents for $\ell_{AA},
\ell$ and $\ell_{AB}$ are still controversial. The two features of
ballistic and diffusive motions result in pentagonal space-time
trajectories of bulk particles (Fig. 1(a)), which allow us to
derive the asymptotic scaling analytically.


We consider a configuration in which $A$ and $B$ species are
randomly distributed on an one dimensional lattice with an equal
initial density, $\rho_A (0)=\rho_B (0)$. A randomly-chosen
particle performs either isotropic or biased random walks
depending on the configurations of neighboring particles. When the
chosen particle is surrounded by two same species neighbors such
as $BAB$, the chosen particle ($A$) performs isotropic random
walks. If two opposite species particles surround a particle such
as $AAB$, the chosen particle ($A$) is constantly driven to the
its opposite species ($B$).

In the region of a length $\ell$, the number of $A$ species is
initially $N_A = \rho_A (0) \ell \pm \sqrt{\rho_A (0)\ell}$ and
the same for $N_B$. After a time $t \sim \ell^z$, particles travel
throughout the whole of the region, and annihilate in pairs. The
residual particle number is the number fluctuation in the region
so we have the relation $N_A \sim \sqrt{\ell}$ or $\rho_A \sim
1/\sqrt{\ell}$ for a given length $\ell$ \cite{monopole,kang}. As
the processes evolve, the system becomes a homogeneous collection
of alternating $A$-rich and $B$-rich domains. To characterize the
structure of segregated domains, we introduce three length scales
as in Ref. \cite{length}. The length of the domain ($\ell$) is
defined as the distance between the first particles of adjacent
opposite species domains \cite{length}. The length $\ell_{AB}$ is
defined as the distance between two adjacent particles of opposite
species, while $\ell_{AA}$($\ell_{BB}$) is the distance between
adjacent $A$($B$) particles in a $A$($B$) domain. These length
scales asymptotically increase in time as
\begin{equation}
\ell \sim t^{1/z}, \;\ell_{AA} \sim t^{1/z_{AA}}, \;\ell_{AB} \sim
t^{1/z_{AB}} \;.
\end{equation}
With the attractive interaction, as a bulk particle inside single
species domains diffuses isotropically until it becomes a boundary
particle, its space-time trajectory is pentagonal as shown in Fig.
1(a). These pentagonal trajectories should be self-similar
(self-affine) fractal structures, because they should have the
scaling symmetry under the scaling transformation $x'=bx$
(space-domain scaling) and $t'=b^z t$ (time-domain scaling) with
$b>1$ due to the power-law behavior in Eq. (1). A typical base
unit of the self-similar pentagonal trajectories of adjacent
opposite domains are schematically depicted in Fig. 1(b). This
base unit allows us to calculate a time $\tau_\ell$ needed to
remove the unit of the space-domain size $\ell$ surrounded by one
scale larger ones. Then the size of the larger unit increases by
$\ell$ during $\tau_\ell$ so we have
\begin{equation}
d \ell /dt \sim \ell /\tau_{\ell} \;,
\end{equation}
which gives the dynamic exponent $z$.
\begin{figure}
\includegraphics[width=7cm]{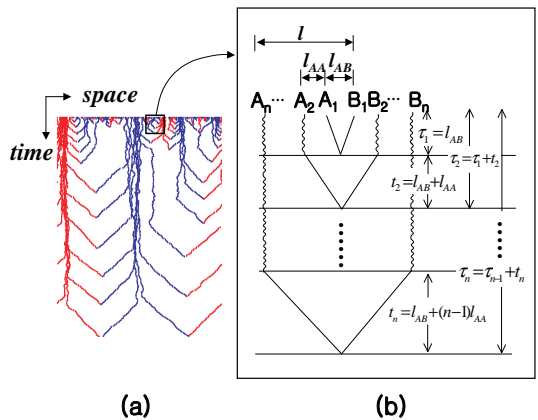}
\caption[0]{\label{pattern} (a) A snapshot of space-time
trajectories of $A+B \rightarrow 0$ with the local attractive
interaction between opposite species. (b) The magnified schematic
space-time trajectories of one pentagonal base configuration of
adjacent opposite species domains. Subscripts \{1,2,...,n\}
indicate the order of the positions of particles from a given
domain boundary. }
\end{figure}

As only boundary particles of each domain have two opposite
species neighbors, the boundary particles ballistically annihilate
due to the attractive interaction. It takes a time $\tau_1 =
\ell_{AB}$ for two boundary particles to annihilate in the base
unit. The second particle from the boundary isotropically diffuses
during time $\tau_1$ until the boundary particles annihilate.
After the time $\tau_1$, the second particle becomes a boundary
particle, and constantly moves to its counterpart during $t_2 =
\ell_{AB} + \ell_{AA}$. So it takes time $\tau_2 = \tau_1 + t_2$
in total for the second particle to annihilate. Similarly, the
$n$th particle from the initial boundary will annihilate after
$\tau_n = \tau_{n-1} + t_n$, where $t_n = \ell_{AB} + (n-1)
\ell_{AA}$ for $n \geq 2$. From the recurrence relation of
$\tau_n$, we find
\begin{equation}
\tau_n = n \ell_{AB} + n(n-1)\ell_{AA} /2 \;.
\end{equation}
As the number of particles in a domain of size $\ell$ is $N_\ell
\sim \sqrt{\ell}$, the time $\tau_\ell$ needed to annihilate the
domain in the base unit is given by
\begin{equation}
\tau_\ell \sim N_\ell \ell_{AB} + {N_\ell}^2 \; \ell_{AA} \; ,
\end{equation}
for $N_{\ell} \gg 1$. In above calculations, we consider the mean
positions of bulk particles, and  neglect the increase of
$\ell_{AA} (t)$ by diffusions during the annihilation of the base
unit. After a smaller unit is completely annihilated by being
embodied into larger one, the remainder of particles redistribute
over the larger unit increased by the size of the annihilated
unit. Hence we approximate $\ell_{AA}(t) = \cdots =
\ell_{AA}(t+\tau_n )= \cdots = \ell_{AA}(t+\tau_\ell )$ during the
annihilation of the unit.

 The scaling of $\ell_{AA}$
is simply $\ell_{AA} \sim \sqrt{\ell}$ from the relation
$\ell_{AA}(t) \sim 1/\rho(t)$. Hence $\ell_{AA}$ scales as
$\ell_{AA} \sim t^{1/2z}$ with $z_{AA} = 2z$. On the other hand,
the change of $\ell_{AB}$ during $\tau_\ell$ should be order of
$N_\ell \; \ell_{AA}$ because of $N_\ell$ successive annihilations
of two opposite particles at boundaries. So we get $d \ell_{AB}/dt
\sim \Delta \ell_{AB} / \tau_{\ell} \sim \ell / t$. With the
scaling of $t \sim \ell^z$, we find that $\ell_{AB}$ follows the
same scaling as $\ell$, i.e. $\ell_{AB} \sim t^{1/z}$ with
$z_{AB}=z$. Finally using the relations, $\ell_{AA} \sim
\sqrt{\ell}$, $N_\ell \sim \sqrt{\ell}$, and $\ell_{AB} \sim
\ell$, we find $\tau_\ell$ from Eq. (4)
\begin{equation}
\tau_\ell \sim \ell^{3/2} \; .
\end{equation}
Substituting Eq. (5) into Eq. (2) and integrating the resultant
equation, we finally arrive the following scaling relation
\begin{equation}
t \sim \ell^{3/2} \;.
\end{equation}
As a result, we find $z = z_{AB} = 3/2$ with the attractive
interaction between opposite species.

From the scaling of $\ell$, $\ell_{AB}$, and $\ell_{AA}$, asymptotic
decays of various densities can be extracted. The density of total
particles ($\rho = 2\rho_A$), the density of pairs of adjacent same
species ($\rho_{AA} = \rho_{BB}$), and the density of pairs of
adjacent opposite species ($\rho_{AB}$) scale as
\begin{equation}
\rho \sim t^{-\alpha} ,\; \rho_{AA} \sim t^{-\alpha_{AA}} ,\;
\rho_{AB} \sim t^{-\alpha_{AB}} \;.
\end{equation}
As $\rho$ is order of $\rho \sim 1/\sqrt{\ell}$, we have $\rho
\sim t^{-1/2z}$ with $\alpha = 1/2z = 1/3$. $\rho_{AA}$ follows
the same scaling of $\rho$ due to $\rho_{AA} \sim 1/\ell_{AA} \sim
1/\sqrt{\ell}$ so $\rho_{AA} \sim t^{-\alpha}$ with $\alpha_{AA} =
\alpha = 1/3$. Finally $\rho_{AB}$ is $\rho_{AB} \sim 1/\ell$,
which leads to $\rho_{AB} \sim t^{-1/z}$ with $\alpha_{AB} = 1/z =
2/3$. Using self-similar structures of space-time trajectories and
scaling arguments for fluctuations of \cite{monopole,kang,length},
we find following exponents for the reactions $A+B \rightarrow 0$
with the local attractive interaction between opposite species
\begin{equation}
\begin{array}{cc}
\alpha =\alpha_{AA} = 1/3\;\;,\;\; \alpha_{AB} =2/3 \;, \\
 z=z_{AB} = 3/2 \;\;,\;\; z_{AA} = 3 \; .
 \end{array}
\end{equation}
Intriguingly and incidentally $\rho(t)$ decays with the same
exponent $1/3$ as that of the uniformly driven hard-core particles
\cite{hardcore,hardcore2}, in which the driven motion of a single
species domain was argued to be described by the noisy Burgers
equation \cite{hardcore}. However for the attractive interaction
case, the $\frac{1}{3}$ decay of $\rho(t)$ comes from the
interplay of isotropic diffusions inside domains and ballistic
annihilations at boundaries. For scalings of inter-domain
distances and others, our results of (8) are completely different
from those of \cite{hardcore2}, where $\ell \sim t^{7/12},
\ell_{AB} \sim t^{3/8}$, and $\ell_{AA} \sim t^{1/3}$. Hence we
conclude that the local attractive interaction between opposite
species forms a new universality class of irreversible $A+B
\rightarrow 0$ reactions.

To see the validity of our analytic results, we now want to report
the simulation results for the model with the attractive
interactions.  With equal initial density of $\rho_A(0) =
\rho_B(0)$, $A$ and $B$ particles distribute randomly on a lattice
of size $L$. In the simulations we consider both hard-core (HC)
particles and the particles without the HC interactions, which we
call the bosonic particles. In the model with HC particles there
can be at most one particle of a given species on a site. In the
bosonic model there can be many particles of the same species on a
site. As we shall see, the simulation results are independent of
the HC interactions.

All the simulations are done on the one-dimensional chains with
the size up to $L = 3 \times 10^6$ and the initial densities are
always taken as $\rho_A(0) = \rho_B(0) = 0.2$. We average
$\rho(t)$, $\rho_{AA}(t)$, and $\rho_{AB}(t)$ up to $10^5$ time
steps over $7200$ independent runs. In Fig.2, we plot the
densities and their effective exponents defined as
\begin{equation}
- \alpha (t)= \ln[\rho(t)/\rho(t/2)]/\ln 2 \;,
\end{equation}
and similarly for others. As you can see in Fig. 2, the data for
HC particles (Fig.2(a)) are almost identical to those for bosonic
particles (Fig.2(b)). While $\alpha_{AB}$ still shows larger
fluctuations for both HC particles and bosonic particles, $\alpha$
and $\alpha_{AA}$ nicely converge to the same value. We estimate
$\alpha = 0.33(1)$, $\alpha_{AA} = 0.33(1)$, and $\alpha_{AB} =
0.68(2)$ for both HC and bosonic models, which agree well with the
predictions (8).

\begin{figure}
\includegraphics[width=7cm]{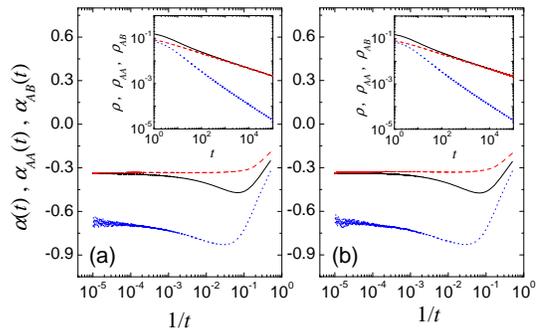}
\caption{\label{rho-hc} densities(inset) and effective exponents
for the densities in the model with hard-core particles (a) and
those in the model with bosonic particle (b). From top to bottom,
each line corresponds to $\alpha_{AA}$, $\alpha$ and
$\alpha_{AB}$. In the inset, the order is $\rho$, $\rho_{AA}$, and
$\rho_{AB}$ respectively. }
\end{figure}

To estimate the dynamic exponent $z$, we measure the densities for
various $L$ from $2^{14}$ up to $2^{18}$. With the scaling
assumption
\begin{equation}
\rho (L,t) \sim t^{-\alpha} f(t/L^z) \;,
\end{equation}
and the estimate $\alpha = 0.333$, we observe the best data
collapse at $z=1.50(2)$ which also agree well with the prediction
of (8) as shown in Fig. 3.
\begin{figure}
\includegraphics[width=7cm]{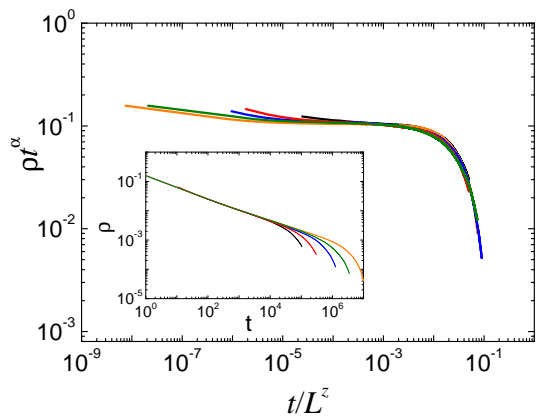}
\caption{\label{scaling-hc} The scaling collapse plot of measured
$\rho(t)$ for Eq. (10). Used sizes  of lattices are $L=
2^{14},...,2^{18}$ and used particles are hard-core particles. We
set $\alpha = 0.333$ and $z=1.5$. The inset shows the raw data for
$\rho(t)$. We do not show the scaling collapse plot for bosonic
particles, which are almost identical to the main plot.}
\end{figure}

For the time dependence of average distances defined in Eq. (1),
we measure $\ell$, $\ell_{AA}$, and $\ell_{AB}$ under the same
measurement conditions as those of densities. The effective
exponents of the distances are defined similarly as in Eq. (9)
except the minus sign. The results for the average distances are
shown in Fig. 4. Here we can also see that the data for HC
particles (Fig. 4(a)) are almost identical to those for bosonic
particles (Fig. 4(b)). From Fig. 4, we estimate $1/z = 0.683(3)$,
$1/z_{AA} = 0.338(3)$, and $1/z_{AB} = 0.66(1)$ respectively for
both HC and bosonic particles.

\begin{figure}
\includegraphics[scale=0.5]{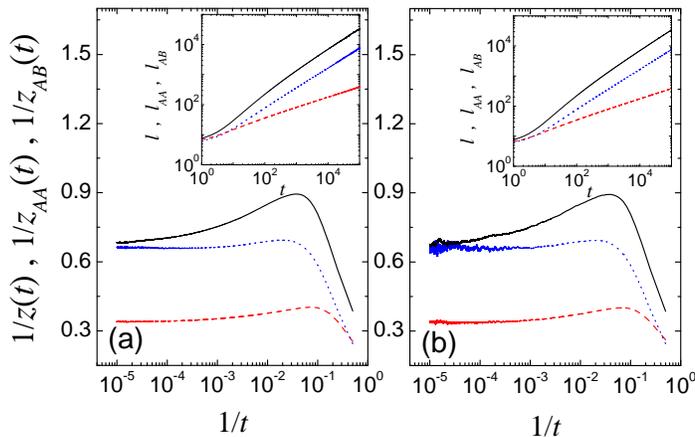}
\caption{\label{distances} The average distances(inset) and
effective exponents for the distances in the model with HC
particles (a) and with bosonic particles (b). From top to bottom,
each line corresponds to $1/z$, $1/z_{AB}$, and $1/z_{AA}$
respectively. In the inset the order of lines is the same as in
the main plot.}
\end{figure}

All the simulation results in Figs. 2-4 numerically confirms the
predictions (8). They also confirms the irrelevance of the HC
interactions in the case of the attractive interaction unlike that
of the uniformly-driven systems \cite{kang,hardcore}. The
isotropic diffusion inside segregated domains leads to Galilean
invariance of domains so the asymptotic scaling behavior is not
affected by the existence of hard-core interaction.

In summary we investigate the anomalous kinetics of $A+B
\rightarrow 0$ reactions with the attractive interaction between
opposite species in one dimension. As reactions proceed, initial
random fluctuations in the particle number develop. As a result,
$A$- and $B$-rich domains appears alternatively, and annihilations
of opposite species take place only at the boundaries of the
closest neighboring domains as in ordinary $A+B \rightarrow 0$
reactions \cite{first,monopole,kang}. However the reactions at
domain boundaries are accelerated by the attractive interaction,
while particles inside domains isotropically diffuses until they
become boundary particles. The interplay of isotropic diffusions
of bulk particles and ballistic annihilations of boundary
particles lead to pentagonal self-similar trajectories which allow
us to analytically calculate the dynamic exponent $z$ and others
as in Eq. (8). We numerically confirm the predictions (8) by means
of Monte Carlo simulations. The results are irrelevant to the
existence of the hard-core interaction between same species
particles.

The anomalous density decay of $t^{-1/3}$ appear to belong to the
same universality class as uniformly driven systems of hard-core
particles, which was argued to be described by noisy Burgers
equations \cite{hardcore}. However our system is not in the same
universality class as the uniformly driven system, because scaling
behaviors of basic distances are different from each other
\cite{hardcore2}. The difference can infer from the underlying
mechanisms. In our model, isotropic diffusions inside domains lead
to Galilean invariance of domains so the hard-core interaction has
no effects on the reactions. Furthermore there is no global bias
to one direction which change the kinetics of hard-core particles.
Only boundary particles feel bias to opposite species, which is
the essential physical factor and distinguishes our model from the
models in Refs. \cite{first,monopole,kang}. We conclude that the
attractive interaction between opposite species is the key feature
of the new universality class characterized by exponents in Eq.
(8), and it is another path to the anomalous density decay of
$t^{-1/3}$.

This work is supported by Korea Research Foundation Grant No.
KRF-2004-015-C00185

\end{document}